\documentclass[%
 reprint,
superscriptaddress,
 amsmath,amssymb,
 aps,
]{revtex4-2}

\usepackage{graphicx}
\usepackage{dcolumn}
\usepackage{bm}
\usepackage{upgreek}

\begin{document}

\title{Light-switchable deposits from evaporating drops containing motile microalgae}

\author{Marius R. Bittermann}
\email{m.r.bittermann@uva.nl}
\affiliation{Van der Waals-Zeeman Institute, IoP, University of Amsterdam, Science Park 904, 1098 XH, Amsterdam, The Netherlands.}
\author{Daniel Bonn}
\affiliation{Van der Waals-Zeeman Institute, IoP, University of Amsterdam, Science Park 904, 1098 XH, Amsterdam, The Netherlands.}
\author{Sander Woutersen}
\affiliation{Van ’t Hoff Institute for Molecular Sciences, University of Amsterdam, Science Park 904, 1098 XH, Amsterdam, The Netherlands.}
\author{Antoine Deblais}
\email{a.deblais@uva.nl}
\affiliation{Van der Waals-Zeeman Institute, IoP, University of Amsterdam, Science Park 904, 1098 XH, Amsterdam, The Netherlands.}


\begin{abstract}
Deposits from evaporating drops have shown to take a variety of shapes, depending on the physicochemical properties of both solute and solvent. Classically, the evaporation of drops of colloidal suspensions leads to the so-called coffee ring effect, caused by radially outward flows. Here we investigate deposits from evaporating drops containing living motile microalgae (\emph{Chlamydomonas reinhardtii}), which are capable of resisting these flows. We show that utilizing their light-sensitivity allows to control the final pattern: adjusting the wavelength and incident angle of the light source enables to force the formation, completely suppress and even direct the spatial structure of algal coffee rings.
\end{abstract}

\maketitle

Evaporation of a drop pinned to a surface induces capillary flows towards its edge, a phenomenon leading to ring-like deposits at the contact line \cite{deegan1997}.
Ways to reverse this so-called coffee ring effect are much sought after, given its undesirability in various technical applications.

Many strategies have been developed to counteract the formation of deposits on drop peripheries such as the introduction of surface tension-driven Marangoni flows \cite{larson2006,seo2017}, capillary forces \cite{weon2010,bonn2015}, prevention of contact line pinning by using hydrophobic \cite{tian2013} or absorbing substrates \cite{Boulogne2015,Boulogne2016}, and even more complex strategies including the drying of frozen drops \cite{Jambon2019}. Most commonly, the evaporation-induced stains are composed of either colloids or macromolecular solutes, in which the transport to the periphery is controlled by advection. Polymer solutions showing low P\'eclet numbers $Pe=Rv/D$, for which diffusive transport, characterized by the diffusion coefficient $D$, dominates over advective flow, defined as the flow velocity $v$ along the drop radius $R$, have shown to mitigate coffee rings in numerical and experimental studies \cite{eales2015,baldwin2011}.

Other systems, which are presumably capable of overcoming advection controlled transport belong to the field of active matter \cite{Marchetti2013}. Active particles, such as biological cells, extract energy from their environment to fuel self-propulsion \cite{bechinger2016}. Drying suspensions of drops of living motile bacteria have shown to lead to disk-like deposits \cite{nelli2007}, to alter coffee rings in colloidal suspensions \cite{andac2019}, to create homogeneous patterns due to the auto-production of biosurfactants \cite{sempels2013} and also to induce complex wetting behaviors in micropillar arrays \cite{susarrey2016,susarrey2018}. While exhibiting a variety of possibilities, the capacity of these active entities to counter act (complex) internal flows in evaporating drops is limited. Due to their weak swimming power, most bacteria cells are subject to internal flows.

\begin{figure}[b]
\centering 
\includegraphics[width=\columnwidth]{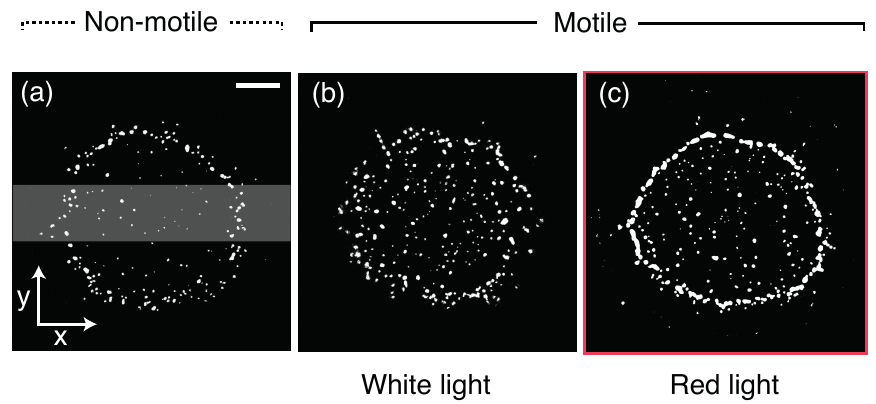}
\caption{\textbf{Phase contrast microscopy images of the deposits left behind drying drops containing \textit{Chlamydomonas reinhardtii} obtained under different light conditions.} (a) Non-motile algae cells leave behind a coffee ring-like deposit. When motile, the final deposit can be either homogeneous, obtained by illumination from above with white light (b), or ring-like, using red light instead (c). Scale bar is $250\,\upmu \mathrm{m}$.}
\label{fig:Fig1}
\end{figure}

Another class of active ``particles'' that potentially surpass this limitation are algae.
\textit{Chlamydomonas reinhardtii} (\textit{C.~R.}) emerged as a model system for the investigation of green microalgae \cite{goldstein2015}, given their well understood interaction with light. Using the synchronous beating of their two flagella, \textit{C.~R.} steer either towards or away from a light source in a process referred to as phototaxis \cite{drescher2010_2,arrieta2017,bruncosmebruny2020,demaleprade2020}, and consequently develop high swimming velocities and large thrust \cite{drescher2010,Guasto2010,Wang2018}. This model organism is even capable of light-induced adhesion to surfaces; It shows strong adhesion to substrates when exposed to blue light \cite{kreis2018}, independently of the chemical composition of the substrate  \cite{kreis2019}. Considering these key characteristics, drops containing \textit{C.~R.} cells offer a promising experimental platform for studying the effect of activity on drying-induced pattern formation.

Using phase contrast microscopy coupled with particle tracking techniques, we show here that by manipulating the light conditions under which evaporation takes place, it is possible to control the final deposits left behind active drops containing \textit{C.~R.} cells. In contrast to bacteria, we find motile algae to counteract advection towards the contact line and undergo mixing within the drops. When dried under white light, \textit{C.~R.} show an increased adhesion on the substrate and an overall lower motility, which leads to a homogeneous final deposit. Using red light instead, increases the motility of the algae and causes them to aggregate at the contact line during the final stages of evaporation, which leads to the formation of a ring-like deposit. This is surprising, given the fact that the algae have sufficient thrust to easily withstand the advective flow that normally causes coffee rings. In fact we will see that, for the algae, the mechanism for the ring-deposit formation is completely different and can be explained in terms of confinement at the contact line related to a critical contact angle. We finally demonstrate that, by adjusting the incident angle of the light source, we can even maneuver the active particles in a desired direction and obtain a crescent-shaped deposit.

To study the influence of motility and wavelength of the incident light on the deposits of drops containing either motile or non-motile  microalgae, $0.3\,\upmu\mathrm{l}$ drops of \emph{Chlamydomonas reinhardtii} (\textit{C.~R.}) suspended in commercial BG-11 medium (from Sigma)  were deposited on glass microscope slides and left to dry at $20^{\circ}\mathrm{C}$ and $50\%$ relative humidity. \textit{C.~R.} are about $10\,\upmu \mathrm{m}$ in diameter and we used a concentration of $\approx 3.7\cdot10^{3} \mathrm{cells}/\upmu \mathrm{l}$.
The aqueous drops exhibited an initial contact angle of $\theta \approx 40^{\circ}$ and the contact line remained pinned during drying (see Materials and methods and Sup.~Fig.~S1) \cite{shahidzadeh-bonn2006, carrier2016}.
We monitored the evaporation process using an inverted phase contrast microscope whilst illuminating the drops from above (See Materials and methods and Sup.~Fig.~S2).
The resulting patterns are summarized in Fig.~\ref{fig:Fig1}. We begin by examining the final patterns left behind a sample containing non-motile algae cells, obtained by heating up the suspension to $\approx 50^\circ$ prior to drop deposition.
This system allows a reference measurement containing inactive particles, as one would get with a regular non-Brownian dispersion of spherical particles.
A dried drop containing these ``passive'' particles left behind a deposit resembling a coffee ring (Fig.~\ref{fig:Fig1}(a)), which we quantified by calculating a rectangular intensity profile (Sup.~Fig.~S3). This is similar to the deposits from drying liquid drops containing inactive colloids, where capillary outward flows carry the solute towards the contact line \cite{deegan1997}.
For motile algae, illumination with regular white light during the evaporation period produces a deposit where dried algae cells are homogeneously distributed (Fig.~\ref{fig:Fig1}(b) and Sup.~Fig.~S3). By implementing a longpass filter, excluding wavelengths below $650\,\mathrm{nm}$ (i.e. illumination with red light), we observed a ring-shaped pattern instead (Fig.~\ref{fig:Fig1}(c) and Sup.~Fig.~S3).
From these observations we surmise that there are two key components to be considered relevant for the final pattern left behind a drying drop containing microalgae: the effect of the algal motility and the influence of the color of the light on the algal motility inside a drying drop.

\begin{figure}[t]
\centering
\includegraphics[width=\linewidth]{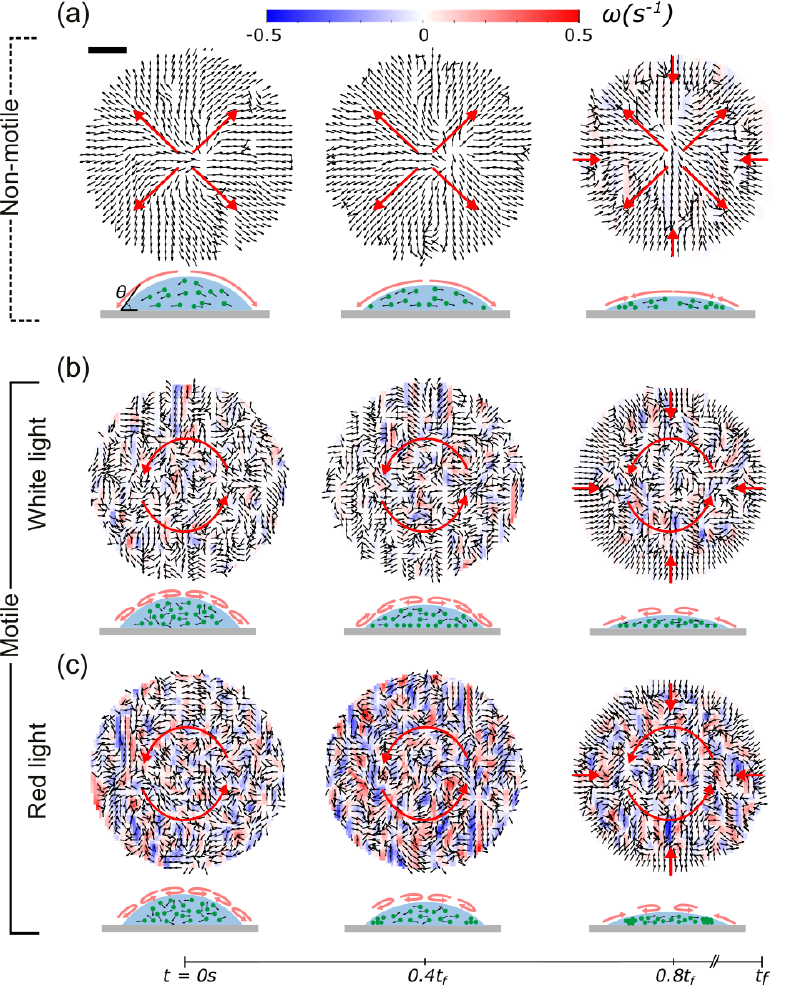}
\caption{\textbf{Velocity fields of the algal drops, as obtained from PTV measurements, show the effect of the motility of the algae on their dynamics during evaporation.} Flow fields are shown for the \textit{non-motile} case (a), and compared to the \textit{motile} cases when illuminated with white light (b) and red light (c). The total drying time $t_{f} \approx 350\,\mathrm{s}$. In the non-motile case, the algae are advected to the contact line (outward flow) and the flow fields are free of any rotations. Motile algae, however,  counteract this outward flow and form multiple vortices. In this case, the living algae ``mix'' within the drop. This effect is stronger when illuminated with red instead of white light.
The displacement towards the drop center at $0.8\,t_{f}$ is due to confinement-induced capillary forces at the contact line.
Corresponding schematics of the side view of the evaporating drops are shown below.
Scale bar is $130\,\upmu \mathrm{m}$}
\label{fig:Fig2}
\end{figure}

To investigate these effects in detail, we used Particle Tracking Velocimetry (PTV), using the algae as particles, to dynamically determine flow fields in the drying drops containing either non-motile or motile cells.
During evaporation, non-motile algae were being advected towards the contact line in a radially outward flow, as shown in Fig.~\ref{fig:Fig2}(a).
In the latter stages of evaporation ($>0.8\,t_{\mathrm{f}}$), we observed an inward motion displacing the non-motile algae towards the drop center.
This was found to be a consequence of capillary forces induced by the confinement of the particles at the drop edge  \cite{weon2010,monteux2011,bonn2015}.
When the algae are motile, the picture provided by PTV is different.
If illuminated with white light from above, the algae form multiple vortices \cite{wioland2013,sumino2012,schaller2010}, and ``mix'' within the drop, as shown in Fig.~\ref{fig:Fig2}(b).
This effect is even more pronounced when illuminated with red light from above (Fig.~\ref{fig:Fig2}(c)), suggesting an increased activity for this case.
For both active cases we found the same inward motion in the later stages of evaporation as for the non-motile case, indicating independence of this effect on the motility. 
The PTV results show that the active algae resist the outward flows observed in the non-motile case and that their activity inside the drops increases when dried under red instead of white light.

\begin{figure}
\centering
\includegraphics[scale=1]{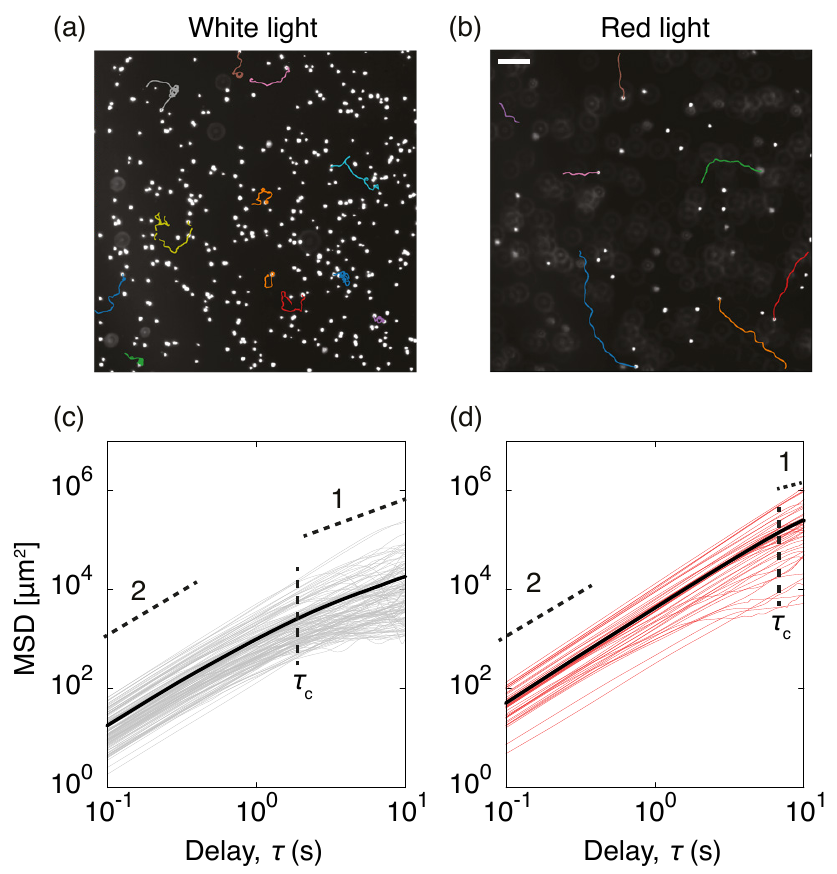}
\caption{\textbf{Effect of the light condition on the dynamics and adhesion of \textit{Chlamydomonas reinhardtii} in evaporating drops.} The colored traces in the phase contrast microscopy snapshots show a few examples of center-of-mass trajectories, for algae fueled with white light (a) and red light (b). In the white light case, the algae remain in the focal plane near the substrate (adhesion), while under red light, the algae are prone to move outside of the focal plane (no adhesion). Scale bar is $50\,\upmu \mathrm{m}$. Corresponding mean square displacements (MSDs) as a function of time for $\approx 50$  algae cells, with the  averaged MSDs (bold lines) for both the white light (c) and the red light case (d). The horizontal dashed lines highlight the initial ballistic regimes (slope 2), which turn diffusive-like on longer timescales (slope 1). The vertical dashed lines show the transition between both regimes marked by the critical delay time $\tau_{\mathrm{c}}$.}
\label{fig:Fig3}
\end{figure}

\begin{figure}
\centering
\includegraphics[width=\columnwidth]{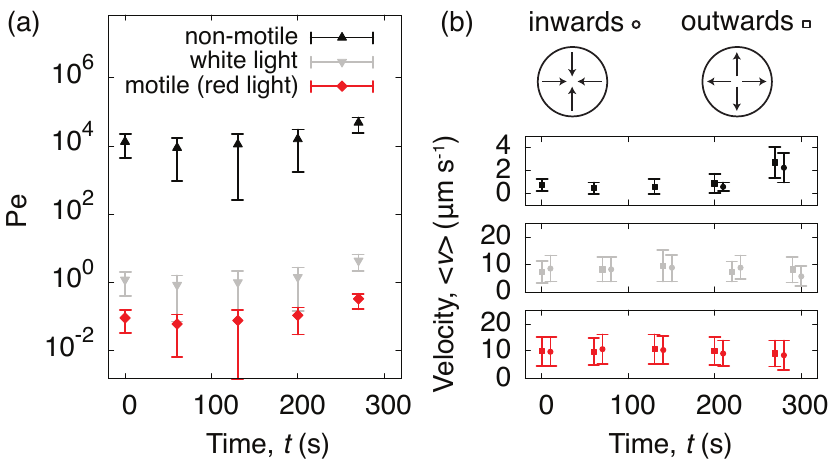}
\caption{\textbf{P\'eclet numbers and velocities of non-motile and motile algae  suspended in evaporating drops.} (a) The active algae resists advection during drop evaporation as shown by the time-dependent P\'eclet numbers, which were calculated by comparing the advection due to the evaporative flux inside the drop (obtained from Fig.~\ref{fig:Fig2}a) against the diffusion coefficients of the algae (obtained from Fig.~\ref{fig:Fig3}c,  d). The P\'eclet numbers decrease with red light compared to white light. (b) The averaged velocities $\langle v \rangle$ are reported for the non-motile case (top), the motile case when illuminated with white light (middle), and the motile case when illuminated with red light (bottom) from above. The velocities remain isotropic for both active cases.}
\label{fig:Fig4}
\end{figure}

To quantify \textit{C.~R.} activity under different light conditions, we tracked the algae directly at the glass surface.
A snapshot of motile algae illuminated with white light from above is presented in Fig.~\ref{fig:Fig3}(a), along with a few selected trajectories. Here the algae move close to the surface, contrary to \textit{C.~R.} under red light (Fig.~\ref{fig:Fig3}(b)), where most algae swim outside the viewing plane.
The increased cell-surface interaction when illuminated with white light has been shown to originate from a light-switchable adhesiveness mediated by contact between the flagella of the algae and the substrate, which is regulated by a blue-light receptor \cite{kreis2018}.
From the trajectories of the algae we calculated their mean square displacements  $\mathrm{MSD}(\tau) = \langle {\lvert \textbf{r}(t+\tau)-\textbf{r}(t) \rvert}^{2} \rangle$, from which we extracted their diffusion coefficients $D$ as $\mathrm{MSD}(\tau)=4D\tau$ for delay times $\tau > \tau_{\mathrm{c}}$, where the critical delay time $\tau_{\mathrm{c}}$ separates an initial ballistic regime from a diffusive-like behavior at longer timescales \cite{brun-cosme-bruny2019}.
The increased cell-adhesiveness when illuminated with white light reflects in the MSDs of the algae. The onset of the ballistic regime at $\tau_{\mathrm{c}}$ is shifted to shorter delay times when illuminated with white light ($\tau_{\mathrm{c}}\approx 2\,\mathrm{s}$, Fig.~\ref{fig:Fig3}(c)) instead of red light ($\tau_{\mathrm{c}}\approx 8\,\mathrm{s}$, Fig.~\ref{fig:Fig3}(d)), and the diffusion coefficient of the white light case, $D\approx 4.6 \times10^2 \upmu \mathrm{m}^2/ \mathrm{s}$, is significantly lower than the one from the red light case, $D\approx 6.2 \times10^3 \upmu \mathrm{m}^2/ \mathrm{s}$. 
To evaluate the competition between the diffusivity of the algae and the rate of advection due to the evaporative flux, we calculate the dimensionless P\'eclet numbers $Pe=R \langle v(t) \rangle / D$, where $R$ is the drop radius and $\langle v(t) \rangle$ the averaged time-dependent velocity due to the evaporative flux, extracted from the flow fields governing the non-motile case in Fig.~\ref{fig:Fig2}(a).
For the non-motile algae, we calculated $D$ using the Stokes-Einstein equation $D=k_{\mathrm{b}}T/(6\pi\eta r)$ with the viscosity of the solvent $\eta=10^{-3}\,\mathrm{Pa\cdot s}$ and the radius of the algae $r=5\,\upmu \mathrm{m}$. The P\'eclet numbers as a function of time (Fig.~\ref{fig:Fig4}(a)) reveal that, regardless of the light condition, the diffusivity of the algae dominates advection ($Pe<1$) and the red light case exhibits P\'eclet numbers an order of magnitude below the white light case, indicating a higher activity. The non-motile algae however show advection dominated evaporation ($Pe>1$), which is typical for evaporating drops of non-Brownian suspensions.
From the PTV measurements and P\'eclet numbers, we inferred that active algae overcome advection and resist being dragged towards the contact line by the radially outward flux. We corroborate this by showing that the velocities of the active algae remain unchanged during evaporation, regardless of whether they face inwards or outwards with respect to the drop center (Fig.~\ref{fig:Fig4}(b)).

\begin{figure}
\centering
\includegraphics[width=\columnwidth]{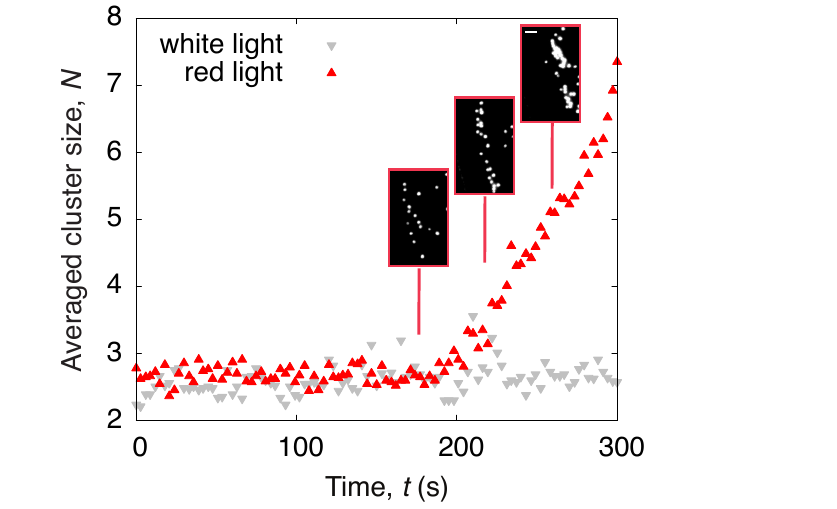}
\caption{\textbf{Time evolution of the averaged number of algae $N$ within a cluster at the drop periphery.} When illuminated with red light from above, clusters appear in the later stages of evaporation. No clusters form when illuminated with white light instead. Phase contrast microscopy snapshots of the red light case show the cluster formation at the contact line (inset). Scale bar is $30\,\upmu \mathrm{m}$.}
\label{fig:Fig5}
\end{figure} 

The question remains why drops containing \textit{C.~R.} leave a ring-like deposit when dried under red light. Density maps obtained from the PTV measurements show that, under red light, active algae accumulate at the contact line in the later stages of evaporation (Sup.~Fig.~S4).
If illuminated with white light from above, the increased cell-substrate adhesiveness, and consequently low diffusivity, prohibits the algae to reach the contact line and the resulting deposit is thus homogeneous.
Considering that the P\'eclet numbers in the red light case remain below 1 during the whole evaporation process (Fig.~\ref{fig:Fig4}(a)), and that the velocities of the algae are isotropic (Fig.~\ref{fig:Fig4}(b)), we can exclude advective flows to contribute to the ring formation.
Zooming into the drop periphery reveals that, from $0.6\,t_{\mathrm{f}}$ onwards, the algae accumulate and form clusters of size $N$ particles (starting at $N=2$).
These clusters grow linearly until the final evaporation time (Fig.~\ref{fig:Fig5}), at which they eventually form the ring-like pattern introduced in Fig.~\ref{fig:Fig1}(c).
We suggest that these clusters form as a consequence of algae getting confined and trapped at the contact line if the drop reaches below a critical contact angle \cite{weon2010,monteux2011,bonn2015,patil2018}, which we verified in a separate experiment involving deposition of drops containing microalgae on a more wetting substrate (Sup.~Fig.~S5).

\begin{figure}
\centering
\includegraphics[width=\columnwidth]{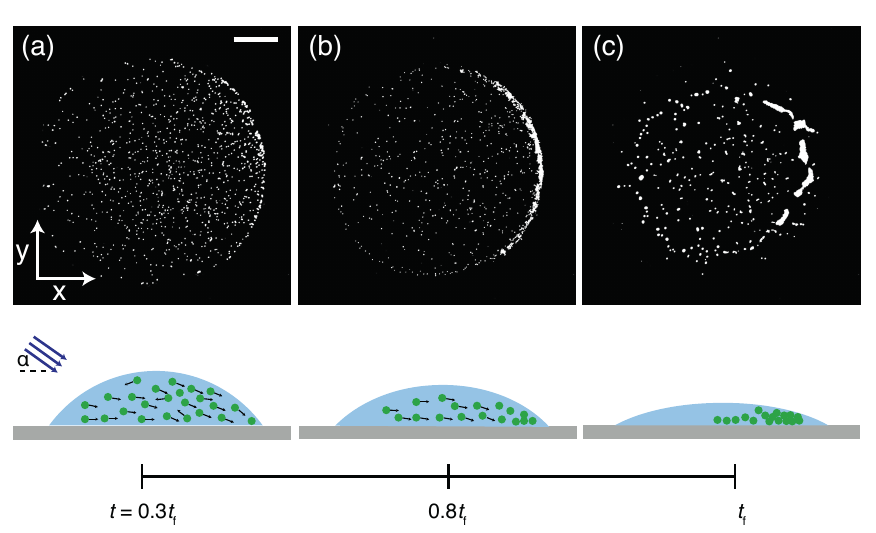}
\caption{\textbf{Taming \textit{Chlamydomonas reinhardtii} to control the final deposit.} Phase contrast microscopy snapshots of a drop containing motile algae cells brought in close contact with a collimated white light source at an angle of $\approx 45^\circ$. When illuminated this way, the algae accumulate at the opposite drop side and end up forming a crescent-shaped final deposit. Corresponding schematics of the side view of the evaporating drops are shown below. Scale bar is $250\,\upmu \mathrm{m}$.}
\label{fig:Fig6}
\end{figure}

In addition to manipulating the deposits left behind drying drops containing \textit{C.~R.} by changing the light color, we here introduce the angle of the incident light as another parameter. This allows to utilize the algal phototaxis \cite{drescher2010_2, arrieta2017, bruncosmebruny2020} to further control the final pattern.
We change from vertical illumination to illumination from the side by placing a collimated white light source at an incident angle of $\approx 45^\circ$ (Sup.~Fig.~S2). In this case we find that the algae accumulate at the opposite drop side in the course of the evaporation (Fig.~\ref{fig:Fig6}). We tentatively attribute this effect to spatial variation of the light intensity inside the drop due to refraction of the light at the air-water interface. If the light hits the droplet at an angle it causes a horizontal intensity gradient and the algal phototaxis in response to this intensity gradient might explain the observed pattern. The collective motion of the algae towards the drop side then results in a crescent-shaped deposit at the final evaporation time $t_{\mathrm{f}}$, highlighting the possibility to manipulate the final deposit and produce on-demand patterns by changing the incident angle of the light source.

To conclude, the final patterns produced by drying drops containing \emph{Chlamydomonas reinhardtii} (\textit{C.~R.}) microalgae vary significantly with their motility, which is tunable by light. If the algae are non-motile, the deposit resembles a coffee ring caused by radially outward capillary flows. Active algae, however, resist this drying induced flow and the deposits can be controlled by adjusting the color of the light source or its angle of incidence under which evaporation takes place. We surmise this variability in the shape of the deposits to originate from different swimming behaviors, which depend on the light condition. This shows that the coffee ring effect in this active matter system can be influenced simply by adjusting parameters of the light source. The findings presented here can be beneficial for a variety of (bio-)technological applications, which are especially important given the recent rise in algal technologies and applications \cite{pierobon2017}.
\newpage
\textit{Acknowledgements}. Authors acknowledge M. van Herk, P. Slot, S. Wilken and C. Sigon from the IBED department of the UvA for providing the algae. 
A.~D. acknowledges the funding from the European Unions’s Horizon 2020 research and innovation program under the individual Marie Skłodowska-Curie fellowship grant agreement number 798455.

\bibliography{references}
\clearpage
\newpage
\renewcommand{\figurename}{SUP. FIG.}
\setcounter{figure}{0}

\section*{APPENDIX A: Materials and methods}

\textbf{Cell cultivation.} \emph{Chlamydomonas reinhardtii}, strain SAG 77.81, were grown axenically in BG-11 medium at $20^{\circ} \mathrm{C}$ at a light intensity of $\approx18\,\upmu\mathrm{mol}/ \mathrm{m^{2}}\mathrm{s}$. Samples from the culture were taken during its logarithmic growing stage and mixed with BG-11 medium to achieve a dilute suspension.\\

\textbf{Substrate and sample preparation.} As a substrate we used a partial-wetting glass slide obtained with a surface silanization technique. Prior to surface silanization, the glass slide was cleaned in a Zepto Electronic Diener plasma cleaner for $60 \mathrm{s}$. Then the glass slide was immersed in a mixture of 1 mL trichloro (octyl)-silane in 99 mL toluene. This procedure introduces a moderate degree of hydrophobicity, which we found to prevent algae cells from sticking to the substrate. The glass slide was cleaned with ethanol and deionized water prior use.\\

\textbf{Experimental settings.} The drying process of the microalgae drops was monitored using an inverted phase contrast microscope at 5x and 10x magnification. A 650 nm longpass filter was added for the experiments involving red light illumination. For the side light measurements, we positioned a collimated white light source at an incident angle of $\approx 45^{\circ}$.\\

\textbf{Tracking.} For the tracking employed in this work, we used the python-based package trackpy \cite{Crocker1996,DanAllan2019}. In the experiments involving the analysis of the whole drops we tracked particles within consecutive frames at 1 frames per second. The position vector is noted \textbf{r}=(x,y) with (x,y) being the in-plane coordinates of the algae. The vorticity \textbf{$\omega$} was calculated from the in-plane velocity \textbf{u}=($u_{x}$,$u_{y}$) as  
\begin{equation}
\omega = \frac{\partial u_{y}}{\partial x} - \frac{\partial u_{x}}{\partial y}
\end{equation}
For the in-depth analysis of the cell-surface interaction we extended the tracking time to 10 seconds recording at 10 frames per second. Within the tracking procedure we always imposed a threshold to filter out non-moving particles.
We extracted the mean square displacement (MSD):
\begin{equation}
\mathrm{MSD}(\tau) = \langle {\lvert \textbf{r}(t+\tau)-\textbf{r}(t) \rvert}^{2} \rangle
\end{equation}
where $\langle\ldots\rangle$ indicates the time average.

\section*{APPENDIX B: Supplementary Information}

\subsection*{Drop shape analysis}
\begin{figure}[t!]
\centering 
\includegraphics[scale=1]{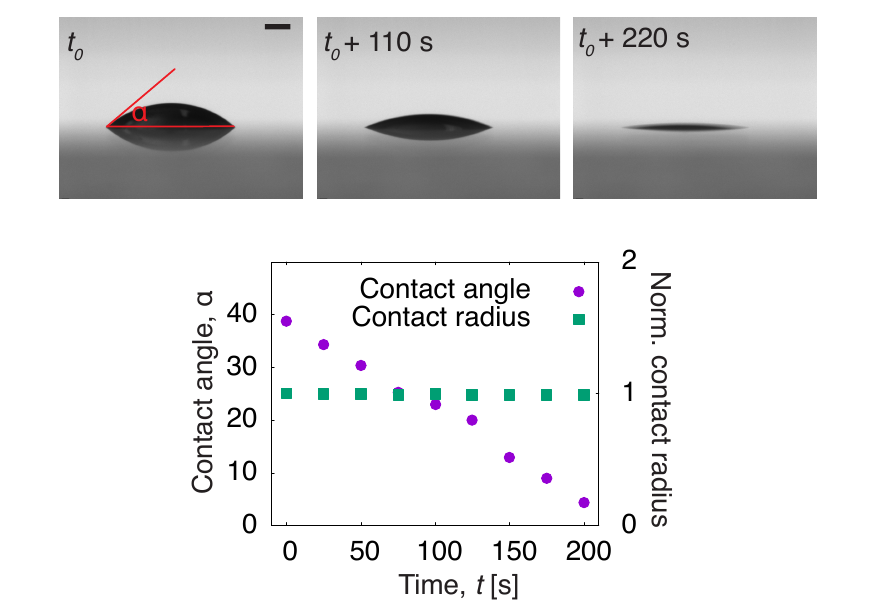}
\caption{Drop shape analysis of a drop containing microalgae. The drop remains pinned during the evaporation process with a steadily decreasing contact angle. Scale bar is $300\,\upmu \mathrm{m}$.}
\label{fig:figS1}
\end{figure}

To measure the time-dependent contact angle of a drop containing \textit{Chlamydomonas reinhardtii} (\textit{C.~R.}) deposited on a glass slide, we used a Drop Shape Analyzer from Krüss. For the measurements, we deposited $0.3\,\upmu\mathrm{l}$ of a drop containing algae on a silanized glass slide and monitored the evaporation every 5 seconds. This allows to extract information on both contact angle and contact radius during the evaporation (Sup.~Fig.~\ref{fig:figS1}).
We observe that the contact angle decreases, while the contact radius remains constant. The drop is pinned to the glass slide during the whole drying process.

\begin{figure}[h!]
\centering 
\includegraphics[width=\columnwidth]{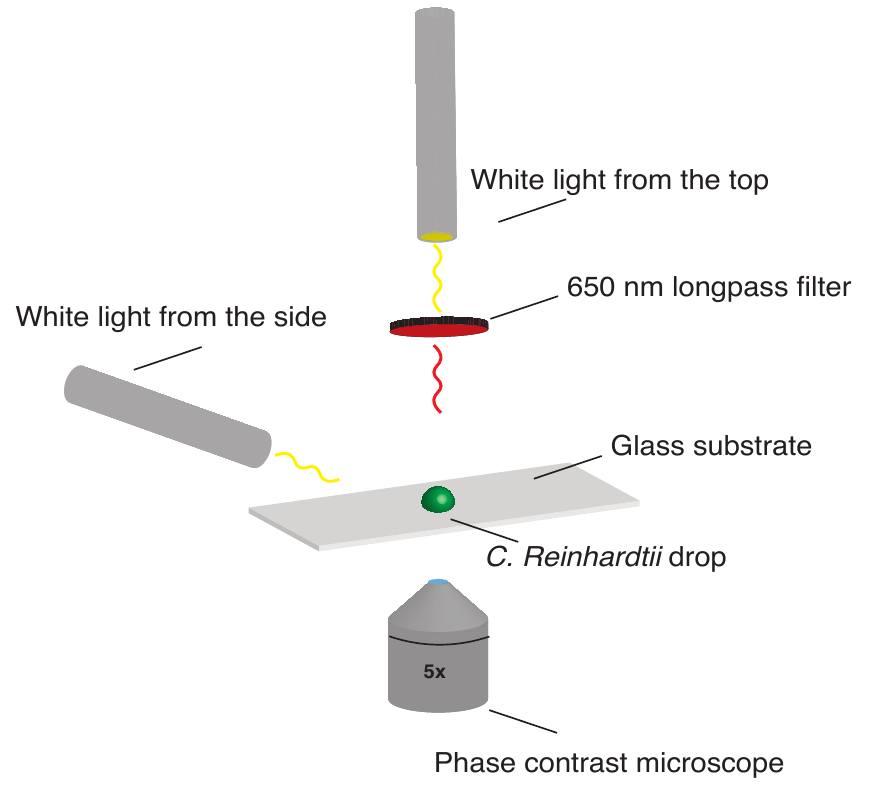}
\caption{Illustration of the experimental setup used in this work.}
\label{fig:figS2}
\end{figure}

\begin{figure}[h!]
\centering 
\includegraphics[width=\columnwidth]{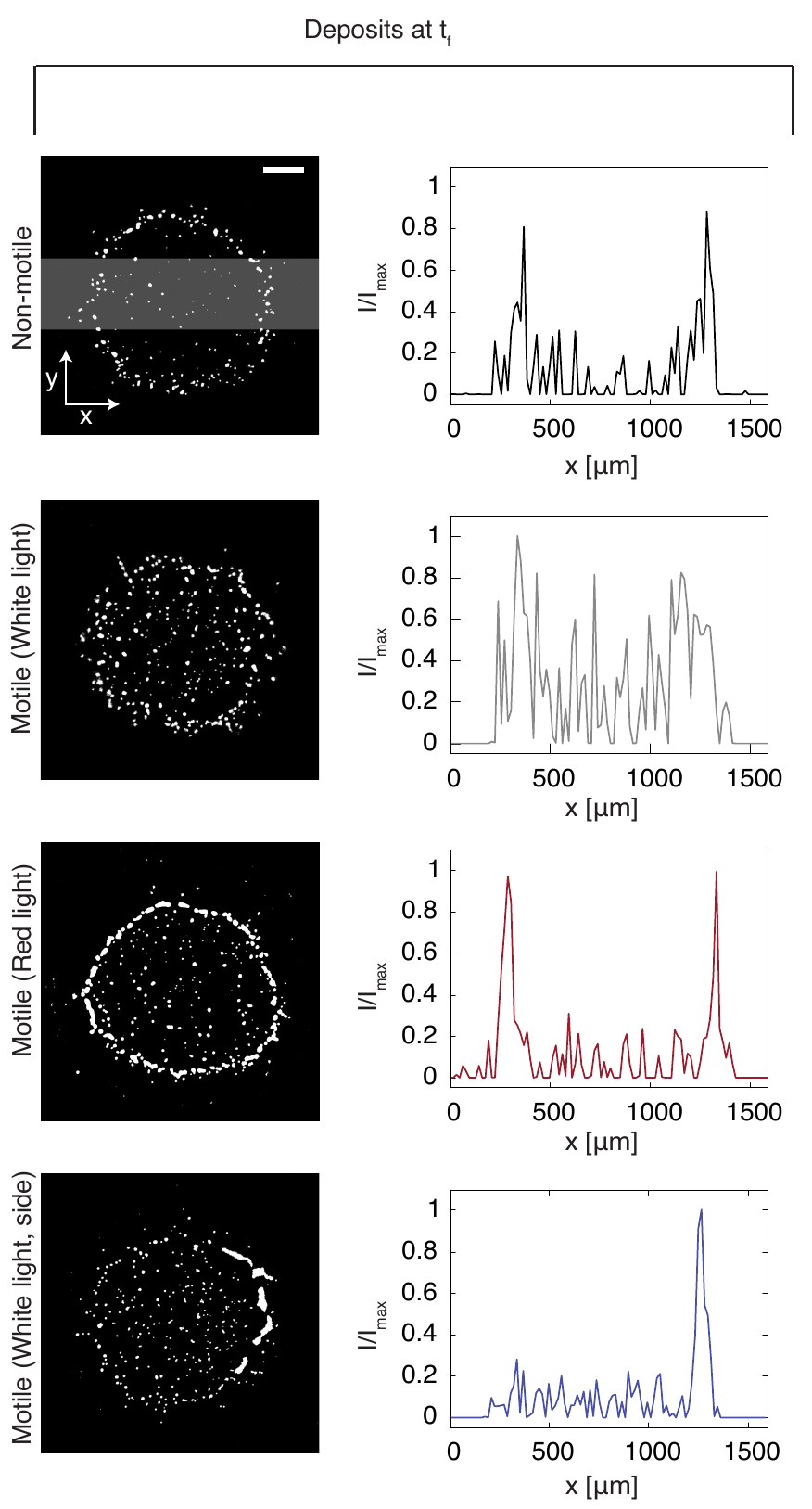}
\caption{Rectangular intensity profiles of the final deposits of drops containing active \textit{C.~R.}. The grey area highlights the analysed section for the corresponding rectangular intensity profiles on the right. Scale bar is $250\,\upmu \mathrm{m}$ }
\label{fig:figS3}
\end{figure}

\subsection*{Experimental setup}
The experimental setup we used is schematically illustrated in Sup.~Fig.~\ref{fig:figS2}. Small drops of motile \textit{C.~R} were deposited on silanized glass slides and left to dry. The evaporation process was monitored using an inverted phase contrast microscope. During evaporation the algae drops were illuminated with either white or red light from above, or white light from the side.

\subsection*{Rectangular intensity profiles}
Rectangular intensity profiles of $\approx1600\times400\,\upmu\mathrm{m}$ were calculated using the image analysis software ImageJ \cite{schneider2012} for all final deposits to emphasize the presence or absence of algal rings (Sup.~Fig.~\ref{fig:figS3}).

\begin{figure}[h]
\centering 
\includegraphics[width=\columnwidth]{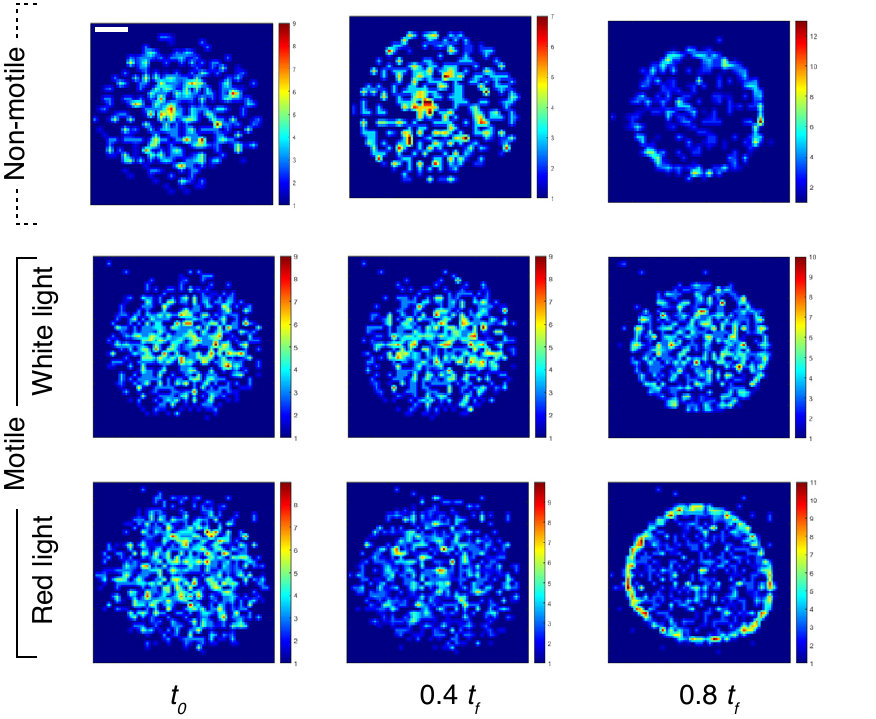}
\caption{Density maps obtained from the PTV measurements show the spatial distribution of \textit{C.~R.} for 3 time steps under different light conditions. Evidently, the algae accumulate at the contact line when illuminated with red light, but remain homogeneously distributed in the white light case. Scale bar is $300\,\upmu \mathrm{m}$. }
\label{fig:figS4}
\end{figure}

\subsection*{Density profiles}
From the PTV measurements we obtained density maps showing the distribution of algae within the drying drops at different time steps under different light conditions (Sup.~Fig.~\ref{fig:figS4}). The data show that active algae accumulate at the contact line during the late stages of evaporation when illuminated with red light from above. This effect is not observed when illuminated with white light instead. 

\subsection*{Ring formation in the red light case}
\begin{figure}[t]
\centering 
\includegraphics[width=\columnwidth]{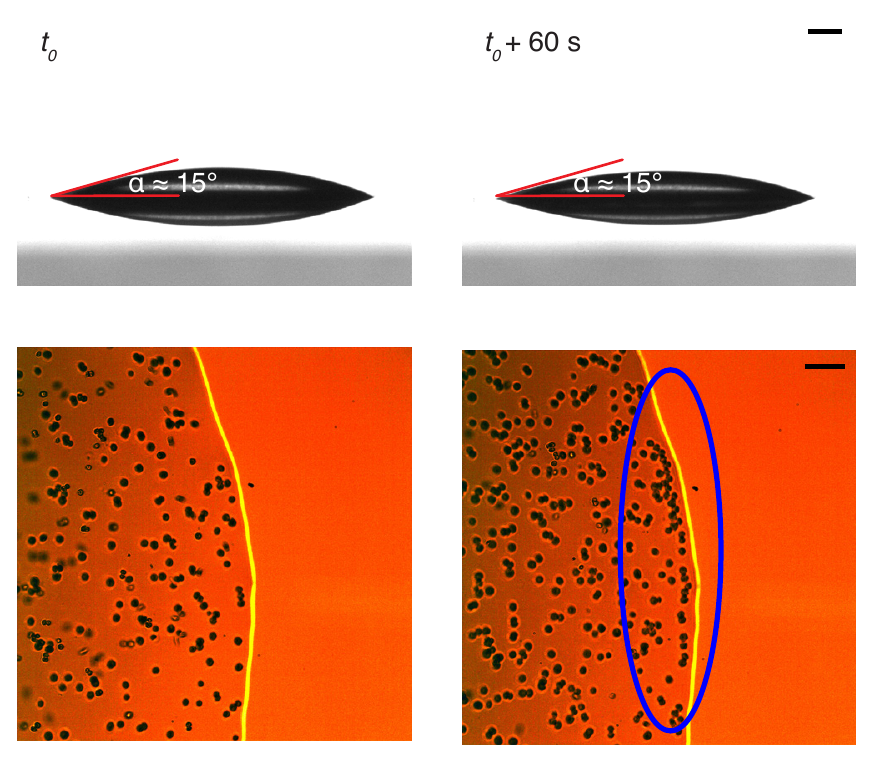}
\caption{Drop shape analysis (top) and bright-field microscopy images (bottom) of the first minute of a large drying drop ($5\,\upmu\mathrm{l}$) on an untreated surface. The small contact angle of $\approx 15^{\circ}$ causes the algae to get trapped at the contact line. Scale bars are $50\,\upmu \mathrm{m}$.}
\label{fig:figS5}
\end{figure}

In the red light case we observed that algae aggregate at the contact line in the later stages of evaporation (Sup.~Fig.~\ref{fig:figS4}, Fig.~5), which eventually led to the ring-like deposit shown in Fig.~1(c).
To verify our hypothesis, that the algae get trapped at the contact line at a critical contact angle $\alpha_{\mathrm{c}}$, we carried out a separate drop shape analysis and bright-field microscopy experiment on a large $5\,\upmu\mathrm{l}$ drop deposited on an untreated glass slide, which exhibits a lower initial contact angle of $\approx 15^{\circ}$ (Fig.~\ref{fig:figS5}) instead of $\approx 40^{\circ}$ (Fig.~\ref{fig:figS1}).
We observed that, although the contact angle remains constant within the time frame of the experiment, the algae aggregate at the contact line. This confirms our assumption that the algae get trapped if the contact angle reaches below a critical contact angle $\alpha_{\mathrm{c}}$.
\vfill\eject

\end{document}